\documentclass[aps,nofootinbib,twocolumn,showpacs,showkeys,tightenlines]{revtex4}
\usepackage{amsmath}
\usepackage{epsf}
\usepackage{epsfig}
\usepackage{axodraw}
\usepackage{amssymb}

\preprint{SNUTP 06-011}
\preprint{KIAS-P06037}
\begin{document}
\title{\Large\bf String MSSM through flipped SU(5) from ${\bf Z}_{12}$
orbifold}
\author{Jihn E.
Kim$^{(a)}$\footnote{jekim@phyp.snu.ac.kr} and Bumseok
Kyae$^{(b)}$\footnote{bkyae@kias.re.kr} }
\address{
$^{(a)}$Department of Physics and Astronomy, and Center for
Theoretical Physics,
Seoul National University,
 Seoul 151-747, Korea, and\\
$^{(b)}$School of Physics, Korea Institute for Advanced Study,
207-43 Cheongryangri-dong, Dongdaemun-gu, Seoul 130-722, Korea}

\begin{abstract}
In a $Z_{12-I}$ orbifold compactification through an intermediate
flipped SU(5), the string MSSM (${\cal S}$MSSM) spectra (three
families, one pair of Higgs doublets, and neutral singlets) are
obtained with the Yukawa coupling structure. The GUT
$\sin^2\theta_W^0=\frac38$, even with exotics in the twisted sector,
can be run to the observed electroweak scale value by mass
parameters of vectorlike exotics near the GUT scale. We also obtain
$R$-parity and doublet-triplet splitting.
\end{abstract}

\maketitle

 \def\Z{{\bf Z}}
 \def\Qem{{$Q_{\rm em}$}}
 \def\pmhalf{{$\pm\frac12$}}
 \def\half{{$\frac12$}}
 \def\SMSSM{${\cal S}$MSSM\ }
 \def\MGUT{{$M_{\rm GUT}$}\ }


Superstring theory is with us for more than 20 years.   Yet there
has not appeared any unique low energy string prediction. Some
relevant phenomenological and cosmological issues are $b-\tau$
unification \cite{BEGN}, unification of gauge couplings at \MGUT
\cite{CouplingU}, the GUT value of $\sin^2\theta_W$ \cite{kim03},
and existence of fractionally charged particles (FCP)
\cite{crypton}. Even though these seem to be most esthetic, any of
these is not an inevitable prediction of string theory. But, if
string theory is valid as a particle physics model, one firm
prediction is that it must lead to the standard model (SM) or the
minimal supersymmetric standard model (MSSM), with some inclusion of
SM singlets.

Obtaining MSSM through simple group GUTs, such as SU(5), SO(10) and
E$_6$, is attractive since it can explain the first three issues
without any FCPs. This simple group GUT scheme needs adjoint
representations for the GUT scale spontaneous symmetry breaking.
However, obtaining adjoint representations from superstring theory
is very difficult if  not impossible
\cite{Schellekens,Tye,ChoiKim05}. So, it has been a recent trend
\cite{nonsimpleGUT} in string compactification to consider
non-simple group GUTs such as flipped SU(5) \cite{Barr},
trinification, and Pati-Salam model. Among these, let us focus on
flipped SU(5).

As pointed out recently \cite{Shafi06}, flipped SU(5) has many nice
features, among which the possibility of GUT symmetry breaking by a
rank-lowering (rank 5$\to$ rank 4)  scalar field is most relevant in
string phenomenology. Representation ${\bf 10}$ achieves this
symmetry breaking pattern. This view was taken in the fermionic
construction by Antoniadis, Ellis, Hagelin, and Nanopoulos (NAHE)
\cite{NAHE}. With the NAHE set of conditions in the fermionic
scheme, there appear \Qem=$\pm\frac12$ particles transforming as
$\bf 4$ or $\overline{\bf 4}$ under a hidden SU(4)$^\prime$, but it
has been known that the resulting cryptons (group-singlet composite
states under the confining gauge group) are integer-charged
\cite{crypton}. On the cosmological side, cryptons with mass in the
range $10^{11}-10^{13}$~GeV have been suggested to present an
important signal via their decay products now in the universe
\cite{Crypcosmo}. Therefore, it is an important phenomenological
issue whether string derived flipped SU(5) models always predict
integer-charged cyptons. Our orbifold construction of flipped SU(5)
show that cryptons are integer-charged if cryptons are indeed
formed. In one construction, we obtain SU(4)$^\prime$
\cite{KimKyae}, but with the hidden sector gauge group and the
spectrum different from those of \cite{crypton}. In this letter, we
consider another construction where the nonabelian part of hidden
sector is $\rm SU(2)^\prime\times SO(10)^\prime$. This is the first
example realizing a flipped SU(5) without SU(4)$^\prime$. Most
probably, the compactification allows color exotics (C-exotics), the
colored particles not having \Qem=$\frac23(\frac{-2}3)$ or
$\frac{-1}{3}(\frac13)$ for color (anti-)triplets, and similarly
defined electromagnetic exotics (E-exotics) and flipped SU(5) GUT
exotics (G-exotics). In the model, there appear G-exotics (including
\Qem=$\pm\frac16$ C-exotics in them) and \Qem=$\pm\frac12$
E-exotics. If lucky, discovery of \Qem=$\pm\frac12$ exotics at low
energy can be a signal to string origin of elementary particles.

For {\bf 10} and $\bf\overline{5}$ representations of SU(5) GUT and
neutral singlet $\nu^c$, one exchanges $u^c\leftrightarrow d^c$ and
$e^c\leftrightarrow \nu^c$ to obtain flipped SU(5). Thus, the matter
representation of flipped SU(5) is represented under SU(5)$\times\rm
U(1)_X$ as
\begin{equation}
\begin{split}
&{\bf 16}_{\rm flip}\equiv {\bf 10}_{\bf 1}+\overline{\bf 5}_{\bf
-3}+{\bf
1}_{\bf 5}=(d^c,q,\nu^c)+(u^c,l)+e^c,\\
& {\bf 5}_{\bf -2} =({D},h_d),\ \overline{\bf 5}_{\bf
2}=(\overline{D},h_u)
\end{split}\label{fliprep}
\end{equation}
where $q$ and $l$ are lepton and quark doublets, $ D$ is
\Qem=$-\frac13$, and $h_{d,u}$ are Higgs doublets giving mass to
$d,u$ quarks. Note that ${\bf 10}_{\bf 1}$, which does not have a
weight at the center of the weight diagram, contains the neutral
component $\nu^c$; thus a VEV of ${\bf 10}_{\bf 1}$ lowers the rank
and breaks flipped SU(5) down to the SM group. The electroweak
hypercharge of (\ref{fliprep}) is given by $ Y=\textstyle\frac15
(X+Y_5)$, where $Y_5={\rm
diag.}(\frac13~\frac13~\frac13~\frac{-1}{2}~\frac{-1}{2})$, and $
X=\textstyle  {\rm diag.}(x~x~x~x~x)$. Spontaneous symmetry breaking
of flipped SU(5) proceeds in two steps via VEVs of $({\bf 10}_{\bf
1}+\overline{\bf 10}_{\bf -1})$ at the GUT scale and $({\bf 5}_{\bf
-2}+\overline{\bf 5}_{\bf 2})$ at the the electroweak scale. In
addition to three ${\bf 16}_{\rm flip}$s, thus flipped SU(5) spectra
must include ${\bf 10}_{\bf 1},\overline{\bf 10}_{\bf -1},{\bf
5}_{\bf -2}$, and $\overline{\bf 5}_{\bf 2}$. These Higgs multiplets
needed for spontaneous symmetry breaking must be vector-like.

 The appearance of fractionally charged particles
is a generic phenomenon in string models, because electromagnetic
charge \Qem\ is not necessarily embedded in an SU(5)-like form.
 In addition, in most standard-like models, the
unification value $\sin^2\theta_W^0$ turns out to be $\le \frac38$.

String MSSM (${\cal S}$MSSM) is defined to be {\it string
compactification with spectra compatible with obtaining} MSSM. It is
not a `string inspired' MSSM but `string derived' MSSM. The particle
contents of \SMSSM are three families of quarks and leptons, one
pair of Higgs doublets without colored scalars, and GUT scale Higgs
bosons responsible for breaking a GUT group down to the SM. Models
with no possibility of fitting $\sin^2\theta_W\simeq 0.23$ at the
electroweak scale are excluded from  ${\cal S}$MSSM. But string
derived standard-like models with $\sin^2\theta_W\simeq 0.23$ at the
electroweak scale are included in  ${\cal S}$MSSM.  ${\cal S}$MSSMs
should possess desirable Yukawa couplings, not obviously excluded by
phenomenology. Earlier standard-like models \cite{iknq} are not
${\cal S}$MSSMs.

The orbifold compactification \cite{DHVW} got much interest due to
its geometric nature and its simplicity in model building. To
construct flipped SU(5) in the orbifold scheme, we adopt $\Z_{12-I}$
twist. If one does not want to introduce any nonabelian group except
SU(5) from the $\rm E_8$ part by one shift $V$, $\Z_{12-I}$ and
$\Z_{12-II}$ are the only possible twists, which can be easily
checked from the Dynkin diagram technique \cite{kacpet}, as
discussed in the expanded version of this letter \cite{KimKyae}. The
$\Z_{12-I}$ shift in the six real (or three complex) internal space
is taken as \cite{ChoiKim05},
\begin{eqnarray}
\textstyle \phi_s= (\frac{5}{12},~\frac{4}{12},~\frac{1}{12})\quad
{\rm with}\quad \phi_s^2=\frac{1}{12}\cdot \frac72.\label{twist}
\end{eqnarray}
This $\Z_{12-I}$ shift is of order 12 for the first and third tori
and is order 3 for the second torus. Internal gauge fields can wind
torus, which are called Wilson lines (WLs).

For the shift and Wilson lines (WLs), we take the following,
\begin{equation}
\begin{split}
&V=\textstyle\left( \frac14~ \frac14~ \frac14~ \frac{1}{4}~ \frac14~
\frac{5}{12}~\frac{6}{12}~  0~ \right)\left( \frac{2}{12}~
\frac{2}{12}~ 0~ 0^5 \right)\\
&a_3=a_4=\textstyle\left( 0^5~0~ \frac{-1}{3}~ \frac{1}{3}~
\right)\left( 0~ 0~\frac{2}{3}~0^5 \right)
\end{split}\label{ShiftVa1}
\end{equation}
which  satisfy all the conditions and give $V^2-\phi_s^2=\frac12$
\cite{DHVW,ChoiKim05}. In twisted sectors wound by  WLs, we must
consider $V_+\equiv V+a_3$ and $V_-\equiv V-a_3$. For twisted
sectors not wound by WLs, we use $V_0\equiv V$. In twisted sectors
wound by  WLs, $V_+^2-\phi_s^2=\frac{5}{6}$ and
$V_-^2-\phi_s^2=\frac{3}{2}$ are used in the multiplicity (${\cal
P}$) calculation.

In the low energy world,  massless modes are important. Massless
modes appear in the untwisted sector $U$ (like bulk modes in extra
dimensional field theory) and in the twisted sectors $T_f$ (like
localized modes at fixed points $f$ in extra dimensional field
theory). We are interested in obtaining massless fields of the $\rm
E_8\times E_8^\prime$ gauge sector of heterotic string. The
masslessness conditions for left and right movers must be satisfied
simultaneously, $M_L^2=M_R^2=0$.

Our discussion will proceed in two steps: firstly find all massless
modes possessing ${\cal N}=1$ supersymmetry and second derive Yukawa
couplings. Supersymmetry in  heterotic string is the symmetry of the
NS and R sectors of right movers. The R sector is represented as
four component half integers $s\equiv(s_0,\tilde s)$. The allowed
values of $s$ determine the chirality (by $s_0$ component) and $U_i$
(by $\tilde s$) of the untwisted sector fields. The supersymmetry
condition for  twist (\ref{twist}) is (See Eq. (3.58) of Ref.
\cite{ChoiKim05}), $\tilde r=\textstyle(-\frac12, \frac12, \frac12)$
so that $\phi_s\cdot \tilde r=0$.

\vskip 0.3cm
 \noindent{\bf Untwisted sector} $U$: The masslessness
condition in $U$ is given by ${P^2 = 2}$. Then, always one can find
a massless condition for right movers. We define $U_i\equiv \tilde
s+\tilde r$ for untwisted sector field nontrivial at the $i$th torus
as $ U_1=(-1,0,0),\quad U_2=(0,1,0),\quad U_3=(0,0,1) $, which
depends on what $\tilde s$ is. The above $U_i$ correspond to the
untwisted $\tilde s=(---),(++-),(+-+)$, respectively.

\paragraph{Gauge group}: The gauge multiplet is found with left
movers by winding momenta $P_{\rm un}$ of length $P_{\rm un}^2=2$,
satisfying $P_{\rm un}\cdot V=0$ and $P_{\rm un}\cdot a_3=0$. So, we
find the unbroken gauge group as
\begin{equation}
[SU(5)\times U(1)_X\times U(1)^3]\times [SU(2)\times SO(10)\times
 U(1)^2]^\prime.
\end{equation}
When flipped SU(5) is broken, three U(1)s spanning over the SU(5)
entry region, satisfying $P_{\rm un}\cdot Q_i=0$,  are considered:
$Q_1=(1~1~1~0~0~0~0~0)()^\prime, Q_2=(0~0~0~1~1~0~0~0)()^\prime,
Q_X=(-2~-2~-2~-2~-2~0~0~0)()^\prime$. In flipped SU(5) with the
above $Q_X$ charge, we obtain $\sin^2\theta_W=\frac{3}{8}$, from Eq.
(10.28) of \cite{ChoiKim05}, $ \sin^2\theta_W={1}/{(1+\sum_i
c_i^2)},$ where $c_i$ are properly defined normalization of U(1)s.

\paragraph{Matter in $U$}: For $P_{\rm un}\cdot V=k/12 $ where
$P_{\rm un}^2=2$ and $k=1,\cdots,11$, also there can appear massless
states. They are interpreted as matter. The ${\cal CPT}$ conjugate
of $k$-bin spectrum appears in $(12-k)$-bin. Thus, in $k=6$ the
${\cal CPT}$ conjugates appear again in $k=6$.

\vskip 0.3cm
 \noindent{\bf Twisted sectors} $T_k$:
The $k^{\rm th}$ twisted sector is distinguished by $kV, k(V+a_3)$,
and $k(V-a_3)$, which are denoted as $\tilde V_{0,+,-}$. For right
movers, the twist in the $k^{\rm th}$ twisted sector is
$\tilde\phi\equiv k\phi_s$. The masslessness conditions for left and
right movers are
\begin{equation}
(P+\tilde V)^2=2(1-c_k)^L-2\tilde N_L,\
(s+\tilde\phi)^2=2(1-c_k)^R\label{condit3}
\end{equation}
where $\tilde N_L$ is the left oscillator number in the $k^{\rm th}$
twisted sector, and $ 2(1-c_k)^L=
\frac{210}{144}(k=1),\frac{216}{144}(k=2),\frac{234}{144}(k=3),
\frac{192}{144}(k=4), \frac{210}{144}(k=5),\frac{216}{144}(k=6) $
for the left movers, and $ 2(1-c_k)^R=
\frac{11}{24}(k=1),\frac{1}{2}(k=2),\frac{5}{8}(k=3),
\frac{1}{3}(k=4), \frac{11}{24} (k=5),  \frac{1}{2}(k=6)$ for the
right movers. 
 It is sufficient to consider
$k=1,2,\cdots,6$ twisted sectors only.

 The multiplicity ${\cal P}$ satisfying the
orbifold condition 
in the $m^{\rm th}$ twisted sector is given by one loop partition
function of string \cite{DHVW,ChoiKim05},
\begin{eqnarray} \label{projnew}
{\cal
P}_m=\frac{1}{N}\sum_{k=0}^{N-1}\chi(\theta^m,\theta^k)~e^{2\pi
ik\Theta_0} ~,
\end{eqnarray}
where $N=12$ for ${\bf Z}_{12}$ orbifolds, and
\begin{eqnarray} \label{phase0}
\begin{split}
\Theta_0 =&
\sum_{j}(N^L_j-N^R_j)\hat{\phi}_j-\frac{m}{2}(V_0^2-\phi_s^2)\\
&+ (P+mV_0)\cdot V_0-(\tilde s+m\phi_s)\cdot{\phi_s} +~{\rm integer}
\end{split}
\end{eqnarray}
where $j$ denotes the coordinates of the 6 dimensional compactified
space running over $\{1,2,3,\bar{1},\bar{2},\bar{3}\}$ in
complexified coordinates, and $\hat{\phi}_j=\phi_{sj}{\rm
sgn}(\tilde\phi_j)$ where ${\rm
sgn}(\tilde\phi_{\overline{j}})=-{\rm sgn}(\tilde\phi_j)$. Here
$\tilde{\phi}_i\equiv k\phi_i$ mod. Z such that
$0<\tilde{\phi}_i\leq 0$, $\tilde{\phi}_{\bar{i}}\equiv -k\phi_i$
mod. Z such that $0<\tilde{\phi}_{\bar{i}}\leq 0$, and
$\phi_{s\bar{j}}\equiv\phi_{sj}$. (If $k\phi_i$ is an integer,
$\tilde{\phi}_j=1$~\cite{KRZnpb,Buchmuller:2004hv}). The
$\chi(\theta^m,\theta^k)$ in Eq.~(\ref{projnew}) denotes the
degenerate factor tabulated in Appendix D of \cite{ChoiKim05}. For
the sectors wound by WLs, $V_\pm$ are used instead of $V_0$. The
untwisted sector $k=0$ and twisted sectors for $k=3,6,9$ are not
affected by WLs since the WL condition, $3a_3=0$, makes it trivial.
So, for $ k=3,6,9$, there is the additional condition, $ (P+kV)\cdot
a_3=0$. For $k\ne 3,6,9$, the multiplicity for each twisted sector
$k(V+m_k a_3)$ is ${\cal P} = \frac13 {\cal P}_k$.

\vskip 0.3cm

\begin{table}[ht]
\begin{center}
\begin{tabular}{c|l|c|c}
\hline
 Sct. &  $P\cdot V;\tilde s\to U_i$& $({\bf SU(5)
})^\chi_{({\bf U(1)_X};U(1)^3;U(1)^{\prime 2})}$ & $\cal P$
\\
\hline  &  $\frac{1}{12};(+-+)\to U_3$& $\overline{\bf 10}_{{\bf
-1};0^5}^{L}$, ${\bf {5}}_{{\bf 3};0^5}^{L}$, ${\bf 1}_{{\bf -5};0^5
}^{L}$ &1

\\
$U$ & $\frac{4}{12};(++-)\to U_2$ &  $\overline{\bf 5}_{{\bf 2};0^5
}^{L}$ &1
\\
$$ &  $\frac{5}{12};(+++)\to U_1$ & ${\bf
10}_{{\bf 1}; 0^5}^{R}$, $\overline{\bf {5}}_{{\bf -3};0^5 }^{R}$,
${\bf 1}_{{\bf 5}; 0^5}^{R}$ &1

\\
 & $\frac{4}{12};(++-)\to U_2$ &  ${\bf 1}_{{\bf 0};0^5
}^{L}$ &1

\\ [0.3em]\hline
\hline Sct. &  $\tilde s\to\chi$& $({\bf SU(5)})^\chi_{({\bf
U(1)_X};U(1)^3;U(1)^{\prime 2})}$ & $\cal P$
\\
\hline
 &  &$\overline{\bf 10}^{L}_{{\bf
-1}; \frac12,0,0;0,0}$   & 4
\\
 $T_6$& $(-\pm -)\to R,L$ &${\bf 10}^{L}_{{\bf
1}; \frac{-1}{2},0,0;0,0}$   & 3
\\

  &  &  ${\bf {5}}^{L}_{{\bf 3};\frac12,0,0;0,0 }$,
 $\overline{\bf {5}}^{L}_{{\bf -3};\frac{-1}{2},0,0;0,0 }$&2\\
 & &${\bf 1}_{{\bf 5};
  \frac{-1}{2},0,0;0,0}^{L}$,${\bf 1}_{{\bf -5};
  \frac{1}{2},0,0;0,0}^{L}$ &2
\\[0.3em]
\hline

 $T_1^0$&  &$\overline{\bf 5}^{L}_{{\bf
\frac{-1}{2}};
\frac{-7}{12},\frac{6}{12},0;\frac{1}{6},\frac{1}{6}}$,
$\overline{\bf 5}^{L}_{{\bf \frac{-1}{2}};
\frac{5}{12},\frac{-6}{12},0;\frac{1}{6},\frac{1}{6}}$& 1\\
$T_1^0$& $(-- -)\to L$& ${\bf 5}^{L}_{{\bf \frac12};
\frac{-1}{12},0,\frac{6}{12};\frac{1}{6},\frac{1}{6}}$ & 1\\
$T_1^-$&  & ${\bf 5}^{L}_{{\bf \frac12};
\frac{-1}{12},\frac{4}{12},\frac{2}{12};\frac{1}{6},\frac{1}{6}}$& 1
\\[0.3em]\hline

 $T_2^0$ & $(---)\to L$ &  ${\bf {5}}^{L}_{{\bf
3};\frac{-1}{6},0,0;0,0 }$, ${\bf 1}_{{\bf -5};
  \frac{-1}{6},0,0;0,0}^{L}$&1
\\[0.3em]\hline
$T_4^0$
 &$(---)\to L$  & ${\bf {5}}^{L}_{{\bf -2};\frac{-1}{3},0,0;0,0 }$& 3
\\
 &  & $\overline{\bf {5}}^{L}_{{\bf 2};\frac{-1}{3},0,0;0,0 }$ & 2
\\[0.3em]\hline

 $T_5^0$& &${\bf 5}^{R}_{{\bf
\frac12}; \frac{7}{12},0,\frac{6}{12};\frac{-1}{6},\frac{-1}{6}}$,
${\bf 5}^{R}_{{\bf \frac{1}{2}};
\frac{-5}{12},0,\frac{-6}{12};\frac{-1}{6},\frac{-1}{6}}$&1\\
 $T_5^0$&  $(-+-)\to R$& $\overline{\bf
5}^{R}_{{\bf -\frac12};
\frac{1}{12},\frac{6}{12},0;\frac{-1}{6},\frac{-1}{6}}$ & 1\\

 $T_5^-$&  &$\overline{\bf 5}^{R}_{{\bf \frac{-1}2};
\frac{1}{12},\frac{2}{12},\frac{4}{12};\frac{-1}{6},\frac{-1}{6}}$
& 1\\[0.3em]
 \hline

\end{tabular}
\end{center}
\caption{A $\Z_{12-I}$ orbifold spectra of the flipped SU(5) sector.
These are $\rm SU(2)^\prime\times SO(10)^\prime$ singlets. In $T_6$,
the ${\cal CPT}$ conjugates appear again in
$T_6$.}\label{tb:spectrum}
\end{table}

Now it is straightforward to calculate the massless modes in each
 sector. For flipped SU(5) fields, the result is summarized in
 Table \ref{tb:spectrum}. In $U$ only, we included the group
 singlet $\bf 1_{0}$. There does not appear any massless matter field
 transforming nontrivially under $\rm SO(10)^\prime$.
 In $T_1$ and $T_5$, there appear G-exotics.
 We have not listed 20 $\rm SU(2)^\prime$ doublets,
 14 E-exotics with \Qem=\pmhalf\
 and 79 \Qem=0 singlets. Five extra U(1)s are also shown, which
 can be broken at high energy scales.

\vskip 0.3cm \noindent{\bf Yukawa couplings}:  Rules for Yukawa
couplings are summarized in \cite{HVyuk,ChoiKim05}.
 For the coupling $U_1^kU_2^lU_3^mT^p$ where $T^p\sim
T_{k_1}^{n_1}T_{k_2}^{n_2}\cdots$ with $p=n_1k_1+n_2k_2+\cdots$, the
Lorentz invariance rule for the ${\bf Z}_{12-I}$ shift is
$(-k+\frac{5}{12}p_1, l+\frac{4}{12}p_2, m+\frac{1}{12}p_3) =
(-1,1,1) {\rm mod.} (n_1,n_2,n_3) $ where $n_1,n_2,$ and $n_3$ for
$\Z_{12-I}$ is $(12,3,12)$, and $p_1,p_2$ and $p_3$ are calculated
using the $H$-momenta. The invariance under a generalized Lorentz
shift $k\phi_s$ in $T_k$ gives  the $H$-momentum conservation. The
$H$-momenta for ${\bf Z}_{12-I}$ twist are
\begin{equation}
\begin{split}
&T_1: \textstyle (\frac{-7}{12}~\frac{4}{12}~\frac{1}{12}),\quad
 T_2: (\frac{-1}{6}~\frac{4}{6}~\frac{1}{6}),\quad
 T_3: (\frac14~0~\frac{-3}{4})\\
&T_4:  \textstyle (\frac{-1}{3}~\frac13~\frac13),\quad
 T_5: (\frac{1}{12}~\frac{-4}{12}~\frac{-7}{12}),\quad
 T_6:(\frac{-1}{2}~0~\frac12).
 \end{split}\label{Hmomenta}
\end{equation}

Consider $T_6$, for example.  If we consider $T_6^2$, we have
$(-1,0,1)$, thus we supply $(0,1,0)$ by $U_2$, and hence $ T_6^2U_2$
is allowed.  For the untwisted sector fields only, there is no $U$
and $UU$ coupling. But cubic couplings can be present in the form
$U_1U_2U_3$. There is no quadratic coupling. All the allowed cubic
terms are, using $T_7$ instead of $T_5$,
\begin{equation}
\begin{split}
&U_1U_2U_3, T_6T_6U_2, T_4T_4T_4, T_2T_4T_6,T_1T_4T_7.
\end{split}\label{Yukawa}
\end{equation}
We tabulated some $H$-quantum numbers for singlet combinations which
enable us to search for higher order terms \cite{KimKyae}. Thus, we
obtain the following.

{(i) \it  MSSM spectrum}: G-exotics   ${\bf 5}_{\bf\frac12}$ and
$\overline{\bf 5}_{\bf-\frac12}$ appear in $T_1$ and $T_5$. They
form vector-like representations and can be removed by $T_1T_4T_7$
couplings with VEVs of singlets in $T_4$. In $T_4$, there are 42
\Qem=0 singlets, which make the removal possible. E-exotics with
\Qem=\pmhalf\ appear in $T_1$ and $T_5$, which can be removed again
by $T_1T_4T_7$ couplings.

There exist $\{{\bf 10}_{\bf 1}, \overline{\bf 10}_{\bf -1}\}$ whose
VEV ($\langle\nu^c\rangle$) breaks the flipped SU(5) down to the SM.
Also, there exist the needed electroweak Higgs fields $h_{d,u}\in
\{{\bf 5}_{\bf -2},\overline{\bf {5}}_{\bf 2}\}$.

In $T_6$ and $T_4$, there appear vectorlike representations $({\bf
10}_{\bf 1}+\overline{\bf 10}_{\bf -1})$s, $({\bf 1}_{\bf 5}+{\bf
1}_{\bf -5})$s, and $({\bf 5}_{\bf -2}+\overline{\bf {5}}_{\bf
2})$s. The vectorlike representations in $T_6$ are removed by
$T_6T_6$ times singlet couplings \cite{KimKyae}. The vectorlike
representation in $T_4$, $2({\bf {5}}_{\bf -2}+\overline{\bf 5}_{\bf
2})^L$, can be removed by $T_4T_4T_4$ couplings where  ${\bf 1}_{\bf
0}$ in  $T_4$ gets a VEV. Thus, $2({\bf 1}_{\bf 5}+{\bf 1}_{\bf
-5})^R, 2(\overline{\bf 5}_{\bf -3}+ {\bf 5}_{\bf 3})^R, 3({\bf
10}_{\bf 1}+ \overline{\bf 10}_{\bf -1})^R$, and $2({\bf {5}}_{\bf
-2} + \overline{\bf 5}_{\bf 2})^L$ are removed at the GUT scale.  In
all these, several singlets with $Q_X=0$ are expected to develop GUT
scale VEVs. Then, we obtain ${\bf 16}_{\rm flip}^R({U_1})+ {\bf
16}_{\rm flip}^R({U_3})+{\bf 5}_{\bf -2}^{R}(U_2)$ from untwisted
sectors, and ${\bf 10}_{\bf 1}^R(T_6), {\bf 1}_{\bf 5}^R(T_2),
\overline{\bf 5}_{\bf -3}^R(T_2), \overline{\bf {5}}_{\bf 2}^R(T_4)$
from twisted sectors. These constitute  three families and one pair
of Higgs quintets. It is interesting to note that the pair of Higgs
quintets, ${\bf 5}_{\bf -2}(U_2)$ \cite{LRno}, and $\overline{\bf
{5}}_{\bf 2}(T_4)$, survives the GUT scale symmetry breaking.
Certainly, it is not allowed to write $M_{\rm GUT}{\bf 5}_{\bf
-2}(U_2)\overline{\bf {5}}_{\bf 2}(T_4)$ since there is no coupling
of the form $U_2T_4$. Also, $S_0 U_2T_4$ with some singlet $S_0$ is
not allowed since (\ref{Yukawa}) does not include such a term. Thus,
the coupling ${\bf 5}_{\bf -2}(U_2)\overline{\bf {5}}_{\bf 2}(T_2)$
must arise from higher order terms, suppressing the Higgs doublet
mass far below the GUT scale. But there exists the coupling of the
type $U_2T_6T_6$ where $U_2= {\bf 5}_{\bf -2}$ and $T_6={\bf
10}_{\bf 1}$ among $2({\bf 10}_{\bf 1}+\overline{\bf 10}_{\bf -1})$
in $T_6$. We require that $\langle{\bf 10}_{\bf
1}\rangle=\langle\overline{\bf 10}_{\bf -1}\rangle\sim M_{\rm GUT}$
by $\langle\nu^c\rangle$. It was shown that this coupling is crucial
in realizing the doublet triplet splitting in flipped SU(5)
\cite{NAHE,Shafi06,Hwang}. We have all the types of fields needed
for the doublet-triplet splitting discussed in \cite{Hwang}. Thus
considering cubic couplings, we obtain the so-called MSSM spectra
with one pair of Higgs doublets. However, the survival hypothesis
\cite{survival} is applicable here also if we include all the higher
order terms. Indeed, there exist higher order terms for ${\bf
5}_{\bf -2}(U_2)\overline{\bf {5}}_{\bf 2}(T_2)$, which however can
be made sufficiently small \cite{KimKyae}.

{(ii) \it $R$-parity}: If we consider cubic couplings of
(\ref{Yukawa}), we can define an $R$-parity in the standard way,
$R=-1$ for matter fermions and $R=+1$ for Higgs bosons.
 A nontrivial parity can be defined as
$R=-1\quad {\rm for\ }\ {\bf 10_{\bf 1}}(U_1), \ {\bf 10_{\bf
1}}(U_3),\ {\overline{\bf {5}}_{\bf -3}}(U_1),\ {\overline{\bf
{5}}_{\bf -3}}(U_3)$, ${\bf 1_5}(U_1)$, ${\bf 1_5}(U_3)$ and $R= +1\
{\rm for\ } {\bf 5}_{\bf -2}(U_2)$. Mixing between the first two
families in the untwisted sector and the third family in $T_5$ and
$T_2^0$ is always possible if VEVs of some neutral singlets are
supposed. Such neutral singlets should preserve all symmetries
relevant at low energies. Even with the mixing terms between
untwisted and twisted matter fields, the $R$-parity relevant in low
energies can still be defined by assigning $R=1$ for the neutral
singlets developing VEVs, and $R=-1\quad {\rm for\ }\ {\bf
10_1}(T_6)$, $\overline{\bf {5}}_{\bf -3}(T_2)$, ${\bf 1_5}(T_2)$.
Then, the allowed Yukawa coupling $T_6T_2T_4$ determines $ R=
+1\quad {\rm for\ }\ \overline{\bf 5}_{\bf 2}(T_4)\ . $ Thus,
$R$-parity can survive down to low energies and hence $R$-parity
conservation for proton longevity is fulfilled in the present model.

{(iii) \it  Quark mixing}: For mixing of fermions, we choose the
quark mixing, since there is one coupling relevant only for one
quark, $T_6T_2T_4$, which is interpreted as the top quark Yukawa
coupling, ${\bf 10_{\bf 1}}\overline{\bf {5}}_{\bf -3}\overline{\bf
{5}}_{\bf 2}.$ Then, $b$ quark  mass arises in terms of $U_2$ Higgs
doublet through $T_6T_6U_2$. But $\tau$ may be placed in the
untwisted sector because the charged lepton in twisted sector obtain
mass at higher order \cite{KimKyae}. If the coupling strength of
$T_6T_2T_4$ and $T_6T_6U_2$ are comparable, a large $\tan\beta$ is
needed to obtain $m_t/m_b\sim 35$. So, two light quark families are
placed in the untwisted sector. These have the cubic couplings of
the form $U_1U_2U_3$, rendering $Q_{\rm em}=-\frac13$ quarks mass.
For $m_b\gg m_s$, it is assumed that this $U^3$ coupling strength is
much smaller than that of $T_6T_6U_2$. At the cubic level, quark
mixing does not appear, but higher order couplings render quarks to
mix \cite{KimKyae}.

{(iv) \it  Fitting $\sin^2\theta_W$}: At the full unification scale,
 $\sin^2\theta_W$ is given by
$\frac38$ \cite{KimKyae}. With a simple assumption that G-exotics
are removed at $M_{GE}$ and E-exotics are removed at $M_{EE}$, we
fit $M_{GE}$ and $M_{EE}$ so that three gauge couplings fall in the
observed values at the electroweak scale by package SOFTSUSY
\cite{SOFTSUSY} which includes two loop effects. Certainly, there
exist solutions for $M_{GE}$ and $M_{EE}$ around \MGUT where \MGUT
is defined to be the converging point of $\alpha_2$ and $\alpha_3$.
Above $M_{\rm GUT}$, at $g_{st}\Lambda_s$ a full unification of
couplings is achieved. For example, a solution set is
\MGUT=$2M_{16}, M_{EE}=6M_{16}, M_{GE}=0.1M_{16},\Lambda_s\simeq
52.7M_{16}$ and $g_s=0.8$ where $M_{16}=10^{16}$ GeV \cite{iwkim}.

In this letter, an  \SMSSM spectrum is obtained in a $\Z_{12-I}$
orbifold construction, Eq. (\ref{ShiftVa1}). The  Yukawa coupling
structure clarified how \SMSSM spectrum is possible.   The
$\sin^2\theta_W^0$ GUT value $\frac38$ is phenomenologically allowed
with exotics removed around the GUT scales. So, these exotics in
general present in `string flipped' SU(5) are not expected to be
light.

 \vskip 1cm \centerline{\bf Acknowledgments}
 We thank K.-S. Choi and I.-W. Kim for useful discussions.
One of us (JEK) thanks KITP where this work was finished. JEK is
supported in part by the KRF grant No. KRF-2005-084-C00001.



\end{document}